\documentclass[aps,onecolumn,prl,superscriptaddress,showpacs,showkeys,floatfix]{revtex4-1}
\usepackage{graphicx,color}% Include figure files
\usepackage{epsfig}
\usepackage{graphicx}
\usepackage{mathrsfs}
\usepackage{color}
\usepackage{bm}
\usepackage{amsmath,amssymb,amsthm,amscd}
\usepackage{bbold}
\usepackage{mdwlist}

 \newcommand{\ketbra}[2]{\vert {#1} \rangle \langle{#2}\vert}

\newcommand{\Tr}{\operatorname{Tr}}

\begin{document}

\title{Detecting lower bounds to quantum channel capacities}

\author{Chiara Macchiavello}
\affiliation{Quit group, Dipartimento di Fisica, 
Universit\`a di Pavia, via A. Bassi 6, 
 I-27100 Pavia, Italy}
\affiliation{Istituto Nazionale di Fisica Nucleare, Gruppo IV, via A. Bassi 6,
  I-27100 Pavia, Italy}

\author{Massimiliano F. Sacchi}
\affiliation{Istituto di Fotonica e Nanotecnologie - CNR, Piazza Leonardo
  da Vinci 32, I-20133, Milano, Italy}
\affiliation{Quit group, Dipartimento di Fisica, 
Universit\`a di Pavia, via A. Bassi 6, 
 I-27100 Pavia, Italy}
	
\date{\today}

\begin{abstract} 

We propose a method to detect lower bounds to quantum capacities of a noisy
quantum communication channel by means of few measurements. The method is 
easily implementable and does not require any knowledge about the channel.
We test its
efficiency by studying its performance for most well known single
qubit noisy channels and for the generalised Pauli channel in arbitrary finite 
dimension. 

\end{abstract}

\maketitle

Noise is unavoidably present in any communication channel. In this case the 
ability of the channel to convey information is lower than in the ideal 
noiseless case and it can be quantified in terms of channel capacities.
Depending on the task to be performed and on the 
resources available, several kinds of capacities 
can be defined. The ability of the channel to convey classical information is
quantified in terms of the classical channel capacity $C$ 
\cite{hol,SW}, defined as the 
maximum 
number of classical bits that can be reliably transmitted per channel use. 
If the sender
and the receiver share unlimited prior entanglement, the capacity of 
transmitting classical information is quantified in terms of the 
entanglement-assisted classical capacity $C_E$ \cite{thapliyal,hol2}.  
The ability of the channel to convey classical information privately is
quantified in terms of the private channel capacity $P$, defined as the 
maximum number of classical bits that can be reliably transmitted per channel 
use in such a way that negligible information can be obtained by a third 
party \cite{devetak}. 
The ability of the channel to convey quantum information is quantified in terms
of the quantum capacity $Q$ \cite{lloyd,barnum},  
defined as the maximum 
number of qubits that can be reliably transmitted per channel use.
\par In many practical situations a complete knowledge of the kind of noise
present along the channel is not available, and sometimes noise 
can be completely
unknown. It is then important to develop efficient means to establish whether
in these situations the channel can still be profitably employed for 
information transmission. A standard method to establish this relies on
quantum process tomography, where a complete reconstruction of the CP map
describing the action of the channel can be achieved, and therefore all
its communication properties can be estimated. This, however, is a 
demanding procedure in terms of the number of different measurement settings
needed, since it scales as $d^4$ for a finite $d$-dimensional quantum system. 
In this Letter we address the situation where we want to gain some information 
on the channel ability to transmit quantum information by employing a smaller 
number of measurements, that scales as $d^2$.
We derive a lower bound on the channel capacities that can be experimentally
accessed with a simple procedure and can be applied to an unknown
quantum communication channel. The efficiency of the method is then studied 
for many examples of single qubit channels, and for the generalised Pauli
channel in arbitrary finite dimension. 

\par In the following we focus on memoryless channels. 
We denote the action of a 
generic quantum channel on a single system as ${\cal E}$ and define 
${\cal E}_N= {\cal E}^{\otimes N}$, where $N$ represents the number of 
channel uses.
The quantum capacity $Q$ measured in qubits per channel use is defined 
as \cite{lloyd,barnum,devetak}
\begin{eqnarray} Q=\lim _{N\to \infty}\frac
{Q_N}{N}\;,\label{qn} 
\end{eqnarray} 
where
%\begin{eqnarray} 
$Q_N = \max
_{\rho } I_c (\rho , {\cal E}_N)$, 
%\;,\label{qnmax} 
%\end{eqnarray} 
and $I_c(\rho , {\cal E}_N)$ denotes the coherent information 
\cite{schumachernielsen}
\begin{eqnarray} I_c(\rho , {\cal E}_N) = S({\cal E}_N (\rho )) - S_e
(\rho, {\cal E}_N)\;.\label{ic} \end{eqnarray} 
In Eq. (\ref{ic}),
$S(\rho )=-\Tr [\rho \log _2 \rho ]$ is the von Neumann entropy, and
$S_e (\rho, {\cal E})$ represents the entropy exchange \cite{schumacher}, i.e.
%\begin{eqnarray} 
$S_e (\rho, {\cal E})= S(({\cal I}_R \otimes {\cal
E})(|\Psi _\rho \rangle \langle \Psi _\rho |)) $,
%\;, \label{se}\end{eqnarray} 
where $|\Psi _\rho \rangle $ is any purification of $\rho $ by means of a 
reference quantum system $R$, namely 
$\rho =\Tr _R [|\Psi _\rho \rangle \langle \Psi _\rho|]$.

\par We will now derive a lower bound for the quantum capacity $Q$ that can be 
easily accessed without requiring full process tomography of the quantum
channel. Since for any complete set of orthogonal projectors $\{\Pi _i\}$ 
one has \cite{NC00} 
$S(\rho )\leq  S(\sum _i \Pi _i \rho \Pi _i)$, then for any orthonormal basis 
$\{ |\Phi _i \rangle \}$ in the tensor product of the reference and the 
system Hilbert spaces one has the following bound to the entropy exchange
\begin{eqnarray}
S_e\left (\rho , {\cal E} \right )\leq H (\vec p)\;,  
\label{se-bound}
\end{eqnarray}
where $H(\vec p)$ denotes the Shannon entropy for the vector of the 
probabilities $\{p_i\}$,  with 
\begin{eqnarray}
p_i = \Tr [({\cal I}_R \otimes {\cal
E})(|\Psi _\rho \rangle \langle \Psi _\rho |) |\Phi _i \rangle\langle\Phi _i|] 
\;.
\label{pimeas}
\end{eqnarray}

From Eq. (\ref{se-bound}) it follows that for any $\rho$ and $\vec p$ one has 
the following chain of bounds 
\begin{eqnarray}
Q \geq Q_1 \geq I_c(\rho , {\cal E}_1)\geq S\left ({\cal E} (\rho )\right )-H(\vec p) 
\equiv Q_{DET}
\;.\label{qvec}
\end{eqnarray}
A lower bound $Q_{DET}$ to the quantum capacity of an unknown channel can 
then be
detected  by the following prescription: prepare a bipartite pure 
state $|\Psi _\rho \rangle $ and send it through the channel 
${\cal I} _R\otimes {\cal E}$, where the unknown channel ${\cal E}$ acts on one
of the two subsystems. Then measure suitable local observables 
on the joint output state to estimate $\vec p$ and 
$S\left ({\cal E} (\rho )\right )$ in order to compute $Q_{DET}$. 
Notice that for a fixed measurement setting, one can infer different vectors 
of probabilities 
pertaining to different sets of orthogonal projectors, as will be clarified 
in the following.   
In principle, one can even adopt an adaptive detection scheme to improve the 
bound (\ref{qvec}) 
by varying the input state $|\Psi _\rho \rangle $. 

We will now be more specific and consider first the case of 
qubit channels. We assume that only the local observables  
$ \sigma _x \otimes \sigma _x $, 
$ \sigma _y \otimes \sigma _y $, and $ \sigma _z 
\otimes \sigma _z $ on the system and on the reference qubits are measured,  
and we want to optimise the bound $Q_{DET}$ given these resources.
First, we  notice that the above measurements allow to measure 
$ \sigma _x$, $ \sigma _y$ and $ \sigma _z$  on the system qubit alone, by ignoring
the statistics of the measurement results on the reference qubit. In this way, 
a complete tomography of the system output state can be performed, and 
therefore the term $S\left ({\cal E} (\rho )\right )$ in Eq. (\ref{qvec})
can be estimated exactly. Furthermore, by denoting the Bell states as
\begin{eqnarray}
&&\!\!\!\!\!| \Phi ^\pm \rangle =\frac {1}{\sqrt 2}(|00 \rangle \pm |11 \rangle )\,, 
%\nonumber \\& & 
\ \  |\Psi ^\pm \rangle =\frac {1}{\sqrt 2}(|01 \rangle \pm |10 \rangle )\,, 
\label{phipsi}
\end{eqnarray}
it can be straightforwardly proved that the local measurement settings 
$\{\sigma _x \otimes \sigma _x,\sigma _y \otimes 
\sigma _y,\sigma _z \otimes \sigma _z \}$
allow to estimate the vector $\vec p$ pertaining to 
the projectors onto the following inequivalent bases
\begin{eqnarray}
B_1= &&\{ a |\Phi ^+ \rangle + b |\Phi ^- \rangle  , 
-b |\Phi ^+ \rangle + a |\Phi ^- \rangle  ,  \nonumber \\& & \
c |\Psi ^+ \rangle + d |\Psi ^- \rangle  , -d  |\Psi ^+ \rangle + c |\Psi ^- \rangle   
\}\;,
\label{b1} 
\\ 
B_2= &&\{ a |\Phi ^+ \rangle + b |\Psi ^+ \rangle  , 
-b |\Phi ^+ \rangle + a |\Psi ^+ \rangle  , \nonumber \\& & 
c |\Phi ^- \rangle + d |\Psi ^- \rangle  , -d  |\Phi ^- \rangle + c |\Psi ^- \rangle  
\}
\;,\label{b2} 
\\ 
B_3= &&\{ a |\Phi ^+ \rangle + i b |\Psi ^- \rangle  , 
i b |\Phi ^+ \rangle + a |\Psi ^- \rangle  , \nonumber \\& & 
c |\Phi ^- \rangle + i d |\Psi ^+ \rangle  , i d  |\Phi ^- \rangle + c |\Psi ^+ \rangle   
\}
\;,\label{b3} 
\end{eqnarray}
with $a,b,c,d$ real and such that $a^2+b^2=c^2+d^2=1$. 
Actually, the measurements corresponding to the above three bases are 
achieved by orthogonal projectors of the form
\begin{eqnarray} 
&&\Pi _{\{a | \Phi ^+\rangle + b| \Phi ^-\rangle\}}=\frac 14 (I\otimes I + 
\sigma _z \otimes \sigma _z  )  
%\\& &  
+\frac {a^2 -b^2}{4}(\sigma _x \otimes \sigma _x - \sigma _y \otimes \sigma _y ) +\frac {ab}{2}(\sigma _z \otimes I + 
I\otimes \sigma _z )\,, 
%\nonumber 
\\ 
&&\Pi _{\{c |\Psi ^+\rangle + d |\Psi ^- \rangle \}}=\frac 14 (I\otimes I - 
\sigma _z \otimes \sigma _z  )  
%\\& &  
+\frac {c^2 -d^2}{4}(\sigma _x \otimes \sigma _x + \sigma _y \otimes \sigma _y ) +\frac {cd}{2}(\sigma _z \otimes I - 
I\otimes \sigma _z )\,,
%\nonumber 
\\ 
&&\Pi _{\{a |\Phi ^+ \rangle + b |\Psi ^+ \rangle \}}=\frac 14 (I\otimes I + 
\sigma _x \otimes \sigma _x  )  
%\\& &  
+\frac {a^2 -b^2}{4}(\sigma _z \otimes \sigma _z - \sigma _y \otimes \sigma _y ) +\frac {ab}{2}(\sigma _x \otimes I + 
I\otimes \sigma _x )\,, 
%\nonumber 
\\ 
&&\Pi _{\{c |\Phi ^- \rangle + d |\Psi ^- \rangle \}}=\frac 14 (I\otimes I - 
\sigma _x \otimes \sigma _x  )  
%\\& &  
+\frac {c^2 -d^2}{4}(\sigma _z \otimes \sigma _z +\sigma _y \otimes \sigma _y  ) -
\frac {cd}{2}(\sigma _x \otimes I - 
I\otimes \sigma _x )\,,
%\nonumber 
\\ 
&&\Pi _{\{a |\Phi ^+ \rangle + i b |\Psi ^- \rangle \}}=\frac 14 (I\otimes I - 
\sigma _y \otimes \sigma _y  )  
%\\& &  
+\frac {a^2 -b^2}{4}(\sigma _z \otimes \sigma _z + \sigma _x \otimes \sigma _x  ) -\frac {ab}{2}(\sigma _y \otimes I - 
I\otimes \sigma _y )\,,
%\nonumber 
\\ 
&&\Pi _{\{c |\Phi ^- \rangle + i d |\Psi ^+ \rangle \}}=\frac 14 (I\otimes I + 
\sigma _y \otimes \sigma _y  )  
%\\& &  
+\frac {c^2 -d^2}{4}(\sigma _z \otimes \sigma _z - \sigma _x \otimes \sigma _x ) +\frac {cd}{2}(\sigma _y \otimes I + 
I\otimes \sigma _y )\,, 
%\nonumber 
\end{eqnarray} 
where $\Pi _{\{a |\Phi ^+\rangle  + b | \Phi ^- \rangle\}}$ denotes the projector 
onto the state $a |\Phi ^+\rangle + b |\Phi ^- \rangle$, and analogously 
for the other projectors.
The probability vector $\vec p$ for each choice of basis 
is then evaluated according to Eq. 
(\ref{pimeas}). The expectation values for terms of the form $\sigma_x\otimes 
I$ (or $I\otimes \sigma _x$) 
can be measured from the outcomes of the observable  $\sigma _x \otimes \sigma _x$ by 
ignoring the measurement results on the second (or first) qubit, and analogously for 
the other similar terms in the above projectors.

Therefore, in order to obtain the tightest bound in (\ref{qvec}) given the 
fixed local measurements $\{\sigma _x \otimes \sigma _x,\sigma _y \otimes 
\sigma _y,\sigma _z \otimes \sigma _z \}$, the Shannon entropy 
$H(\vec p)$ will be minimised as a function of the bases (\ref{b1}-\ref{b3}), 
by varying the coefficients $a,b,c,d$ over the three sets. 
In an experimental scenario, after collecting the outcomes of the measurements 
$\{\sigma _x \otimes \sigma _x,\sigma _y \otimes 
\sigma _y,\sigma _z \otimes \sigma _z \}$, this optimisation 
step corresponds to classical processing of the measurement outcomes.

Our procedure can be generalised for arbitrary finite dimension $d$. 
For simplicity, we will now consider a fixed input maximally entangled state 
$|\Psi \rangle = \frac {1}{\sqrt d}\sum _{i=0}^{d-1} |ii \rangle$, where $d$
is the finite dimension of each subsystem, and a Bell basis 
\begin{eqnarray}
|\Phi _i \rangle = (I_R \otimes U_i ) |\Psi \rangle \;,
\qquad i=0,1,\cdots ,d^2-1 \,,\label{bellui}
\end{eqnarray}
with $U_i$ unitary and  $\Tr [U_j ^\dag  U_i]= d\, \delta _{ij}$.
The detectable bound takes the form
\begin{eqnarray}
Q_{DET}=S\left ({\cal E} \left ( \frac I d \right )\right )
- H(\vec p)
\;,\label{qvecI}
\end{eqnarray}
with 
%\begin{eqnarray}
$p_i = \frac {1}{d^2} \sum _j |\Tr [U_i^\dag A_j]|^2$, 
%\;,\label{pis}\end{eqnarray}
where $\{A_ j \}$ denotes the Kraus operators of the channel 
${\cal E} (\rho) =\sum _j A_j \rho A_j ^\dag$.  
The bound of Eq. (\ref{qvecI}) in this case can be detected by measuring 
$d^2-1$ observables via a local setting and 
classical processing of the measurement outcomes. 
Actually, a set of generalised Bell projectors can be written as follows 
\cite{bellob}
\begin{eqnarray}
\ketbra{\Phi ^{mn}}{\Phi ^{mn}}=\frac 1 d \sum _{p,q=0} ^{d-1} 
e^{\frac {2\pi i}{d}(np-mp) }U_{pq} \otimes U^*_{pq} 
\,,\label{umnbel}
\end{eqnarray}
where $m,n=0,1,\cdots, d-1$, and $U_{mn}$ represents the unitary operator 
$U_{mn}=\sum _{k=0}^{d-1} e^{\frac{2\pi i}{d} km} |k \rangle 
\langle (k + n)\!\!\!\mod d |$.
%\begin{eqnarray}
%U_{mn}=\sum _{k=0}^{d-1} e^{\frac{2\pi i}{d} km} |k \rangle 
%\langle (k + n)\!\!\!\!\!
%\mod d |\;.\label{umn}
%\end{eqnarray}
Hence, a set of  measurements on the eigenstates of $U_{mn} \otimes U^*_{mn}$ 
allows to estimate $Q_{DET}$ in Eq. (\ref{qvecI}).   

As mentioned above, the advantage of this procedure is to require $d^2-1$
measurement settings with respect to a complete process tomography, 
where $d^4 -1$ observables have to be measured. 

We want to point out that all detectable bounds we are providing
also give lower bounds to the private information $P$,  
since $P \geq Q_1$ \cite{NC00}. 
Moreover, we can also derive a detectable lower bound to the 
entanglement-assisted classical capacity. 
Actually, this is defined as $C_E = \max_{\rho } I(\rho , {\cal E}_1)$,
%\;,\label{C_E} 
%\end{eqnarray} 
 where 
%\begin{eqnarray} 
$I(\rho , {\cal E}_1) = S(\rho ) + I_c
(\rho, {\cal E}_1)$. 
%\;.\label{i} 
%\end{eqnarray} 
By considering the procedure outlined above, where a maximally entangled
state $|\Psi \rangle$ is considered as input, we then have the lower bound
%\begin{eqnarray} 
$C_E \geq \log_2 d+ Q_{DET}$. 
%;.\label{C_E-lb} 
%\end{eqnarray} 
\par In the following we will analyse explicit examples of noisy quantum 
channels that are typically encountered in practical implementations, and
analyse in detail the efficiency of the proposed procedure. 
We will consider in particular the most well known channels for qubits, i.e.
the dephasing, the depolarising and the more general Pauli channel, the erasure
and the amplitude damping channel. The Pauli channels are also generalised to
arbitrary dimension. We finally consider a family of qubit 
channels with two Kraus operators. We will always consider 
input states with maximal entanglement between system and reference.

A dephasing channel for qubits with unknown probability $p$ can be written 
as 
%\begin{eqnarray}
${\cal E}(\rho )= \left (1-\frac p 2 \right ) 
\rho + \frac p 2 \sigma _z \rho \sigma _z
\;.$ 
%\end{eqnarray}
 Since it is a degradable channel, 
its quantum capacity coincides with the one-shot single-letter quantum 
capacity $Q_1$, and one has   
%\begin{eqnarray}
$Q=Q_1= 1 - H_2 \left( \frac p2 \right )$, 
where $H_2(x) \equiv  -x \log_2 x -(1-x)\log _2 (1-x)$
is the binary Shannon entropy.
%\;.\label{qcdephase}
%\end{eqnarray}
More generally,  we can consider the  channel 
%\begin{eqnarray}
${\cal E}(\rho )=  
\left (1-\frac p 2 \right ) 
\rho + \frac p 2 U \rho U ^\dag $,
%\;, \end{eqnarray}
for any dimension $d$, with $U$ as unitary and traceless operator. 
The von Neumann entropy of the output state 
${\cal E}(\frac Id)=\frac Id$ is given by $S\left ({\cal E}\left (\frac Id \right )\right )
=\log_2 d$. Using the Bell basis (\ref{umnbel}) one obtains 
the detectable bound 
\begin{eqnarray}
Q_{DET}= \log_2 d  - H_2 \left (\frac p2 \right )\;.\label{bounderase}
\end{eqnarray}
This detectable quantum capacity clearly coincides with the quantum channel capacity for $d=2$. 

\par
\begin{figure}[htb]
  \includegraphics[scale=0.8]{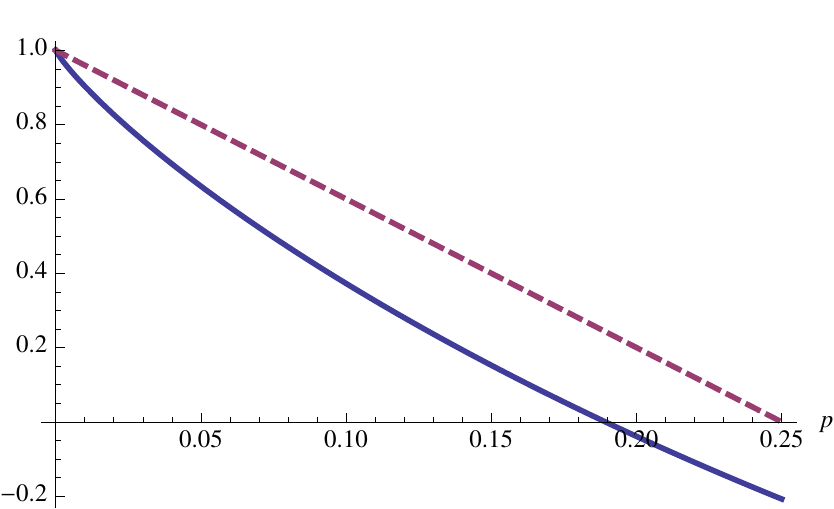}
  \caption{Detectable quantum capacity (thick line) for the depolarising 
channel with error probability $p$ [which coincides with the hashing bound 
(\ref{hash})] versus $p$. 
The dashed line represents the known upper bound $Q \leq 1 - 4p$.}
  \label{fig:depo}
\end{figure}

 The depolarising channel with probability $p$ for qubits is given by \cite{NC00}
%\begin{eqnarray}
 ${\cal E}(\rho )=(1-p)\rho +\frac p3 \sum _{i=x,y,z}\! \sigma _i \rho \sigma _i $.  
%\nonumber \\& =& 
%\left ( 1 - \frac {4p}{3}  \right )\rho  + \frac {2p}{3} I \;. 
%\end{eqnarray}
The quantum capacity is still unknown, although one has the upper bound \cite{ssw} 
$Q \leq 1 - 4p$, 
thus showing that $Q=0$ for $p\geq \frac 14$. 
On the other hand, by random coding the following hashing bound \cite{hashing}
has been proved
\begin{eqnarray}
Q \geq 1- H_2 (p) - p\log _2 3\;.\label{hash}
\end{eqnarray}
This lower bound coincides with our detectable bound of Eq. (\ref{qvecI}). 
%with $\vec p= \{ 1-p, p/3,p/3,p/3\}$. 
In Fig. \ref{fig:depo} we plot the lower bound (\ref{hash}), 
along with the upper bound $Q \leq 1 - 4p$, versus the probability $p$. As we can see, 
our procedure allows to 
detect $Q(p)\neq 0$ as long as $p <  0.1892 $.
\par In arbitrary dimension $d$ the depolarising channel takes the form
\begin{eqnarray}
{\cal E}(\rho )=
\left ( 1 - p \frac {d^2}{d^2-1}  \right )\rho  + p 
\frac {d^2}{d^2 -1}  \frac Id \;. 
\end{eqnarray}
Hence, the detectable bound is simply generalised to 
\begin{eqnarray}
Q \geq Q_{DET}=\log _2 d - H_2 (p) - p\log _2 (d^2 -1)\;,\label{hashd}
\end{eqnarray}
and can be detected by estimating $\vec p$ pertaining to the Bell projectors 
(\ref{umnbel}). 

\par Similarly to the depolarising channel, for a generic Pauli channel 
${\cal E}(\rho )=\sum _{i=0}^3 p_i \sigma _i \rho \sigma _i $
the hashing bound \cite{hashing} provides a lower bound to the quantum 
capacity $Q\geq 1- H(\vec p)$, 
which coincides with our detectable bound (\ref{qvecI}) by 
using a maximally entangled input state and estimating $\vec p$ for the Bell basis. 
In dimension $d$ one can consider the generalised channel 
%\begin{eqnarray}
 ${\cal E}(\rho )=\sum _{m,n=0}^{d-1} p_{mn} U_{mn} \rho U^{\dag }_{mn}\;,$  
%\end{eqnarray}
%in terms of the unitary operators (\ref{umn}), 
and one has 
\begin{eqnarray}
Q \geq Q_{DET}=\log _2 d - H(\vec p) \;, 
\end{eqnarray}
$\vec p$ being now the $d^2$-dimensional vector of probabilities $p_{mn}$ pertaining to the  
generalised Bell projectors in Eq. (\ref{umnbel}). 

\par We consider now an erasure channel \cite{eras1,eras2} with erasure 
probability $p$, defined as
\begin{eqnarray}
{\cal E}(\rho )= (1-p) \rho \oplus p |e \rangle \langle e | \Tr [\rho ] \;, 
\end{eqnarray}
where $|e \rangle $ denotes the erasure flag which is orthogonal to the system Hilbert space. 
Since it is a degradable channel, its quantum capacity coincides with the one-shot single-letter 
quantum capacity $Q_1$, and one has  \cite{eras2} 
\begin{eqnarray}
Q=Q_1=(1-2p)\log _2 d
\;,\label{qcerase}
\end{eqnarray}
for $p\leq \frac 12$, and $Q=0$ for $p\geq \frac 12$.
The output of any  maximally entangled state $|\Psi  \rangle $ can be written as 
\begin{eqnarray}
\!\!\!{\cal E} \left ( |\Psi  \rangle \langle \Psi  |\right )= 
(1-p)|\Psi  \rangle \langle \Psi  |  \oplus \frac p d 
(I_R \otimes |e \rangle \langle e |) 
\,. 
\end{eqnarray}
A basis constructed by the union of the 
projectors on $|i \rangle \otimes |e \rangle  $ (with $i=0,1,\cdots, d-1$) and Bell projectors 
(where one of them corresponds to  $|\Psi  \rangle \langle \Psi  |$) 
gives a vector of probability (\ref{pimeas}) with $d$ elements equal to
$p/d$ and one element $(1-p)$, while all other terms are vanishing. We
then have $H(\vec p)=H_2(p)+ p \log _2 d$. The von Neumann entropy of the reduced output state 
${\cal E}\left (\frac Id \right )
= (1-p) \frac Id \oplus p 
|e \rangle \langle e | $ is given by 
%\begin{eqnarray}
$S \left ({\cal E} \left ( \frac Id \right )
\right )
=H_2 (p)+(1-p)\log_2 d$.
%\;. 
%\end{eqnarray}
 It then follows that the detectable bound $Q_{DET}$ for the erasure channel 
coincides with $Q$ in Eq. (\ref{qcerase}).       

\par The amplitude damping channel for qubits has the form \cite{NC00}
\begin{eqnarray}
{\cal E}(\rho )= A_0 \rho A_0^\dag + A_1  \rho A_1^\dag \;, 
\end{eqnarray}
where $A_0= |0 \rangle \langle 0| + \sqrt {1- \gamma }|1 \rangle \langle 1|$ and 
$A_1= \sqrt \gamma |0 \rangle \langle 1|$. 
Since it is a degradable channel \cite{gf}, its quantum capacity coincides with the one-shot 
single-letter 
quantum capacity $Q_1$, and one has   
\begin{eqnarray}
Q=Q_1= \max _{q\in [0,1]} H_2((1 - \gamma  )q) - H_2(\gamma q) 
\;,\label{qcdamp}
\end{eqnarray}
for $ \gamma   \leq \frac 12$, and $Q=0$ for $ \gamma \geq \frac 12$. 
For an input Bell state $|\Phi ^+ \rangle $ the output is given by 
\begin{eqnarray}
&&{\cal I}_R\otimes {\cal E} (|\Phi ^+ \rangle \langle \Phi ^+|) = 
\frac 14 (1+\sqrt{1-\gamma })^2 |\Phi ^+ \rangle \langle \Phi ^+|  \label{outdamp}\\& &+ 
\frac 14 (1- \sqrt{1-\gamma })^2 |\Phi ^- \rangle \langle \Phi ^-|  
%\nonumber \\& & 
+\frac \gamma 4 (|\Phi ^+ \rangle \langle \Phi ^-|+|\Phi ^- \rangle \langle \Phi ^+|)
\nonumber \\&  &  +\frac \gamma 4 
(|\Psi ^+ \rangle \langle \Psi ^+| + |\Psi ^- \rangle \langle \Psi ^-|
-|\Psi ^+ \rangle \langle \Psi ^-| - |\Psi ^- \rangle \langle \Psi ^+|)
\;.\nonumber 
\end{eqnarray}

\begin{figure}[htb]
  \includegraphics[scale=0.8]{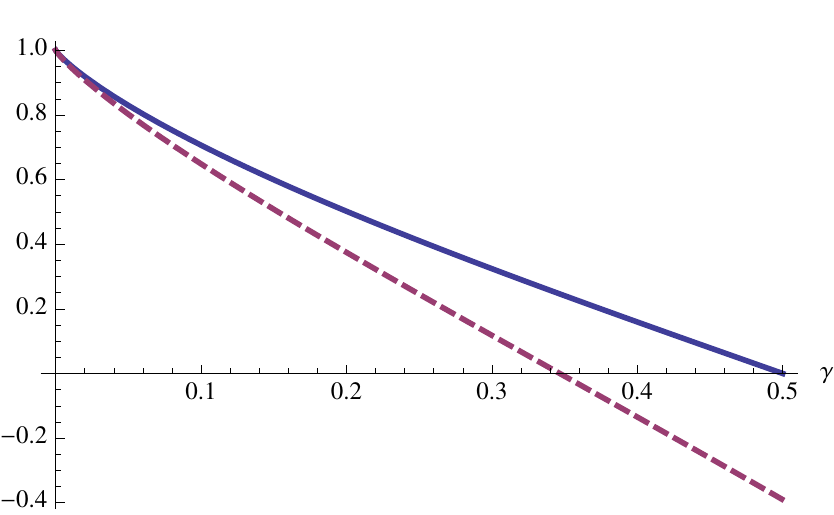}
  \caption{Amplitude damping channel with parameter $\gamma $: 
detected quantum capacity with maximally entangled input and estimation of $\vec p$ for the 
eigenstates of (\ref{outdamp}) and for the Bell basis
(solid and dashed line, respectively).} 
  \label{fig:damp2}
\end{figure}

The reduced output state is given by ${\cal E}\left (\frac I2\right )= 
\frac 12 (I +\gamma \sigma _z)$, 
%\frac{1+\gamma }{2} 
%|0 \rangle \langle 0| +\frac{1- \gamma }{2} 
%|1 \rangle \langle 1|$, 
hence  it has von Neumann entropy 
%\begin{eqnarray}
$S\left ({\cal E}\left (\frac Id \right )\right )
=H_2 \left(\frac {1- \gamma }{2}\right )$. 
%\;. 
%\end{eqnarray}
By performing the local measurement of $ \sigma _x 
\otimes \sigma _x $, $ \sigma _y \otimes \sigma _y $, and $ \sigma _z 
\otimes \sigma _z $ and optimising $\vec p$ over the bases 
(\ref{b1}-\ref{b3}), one can detect the bound 
\begin{eqnarray}
Q\geq Q_{DET}&=& H_2 \left(\frac {1- \gamma }{2}\right )- H(\vec p)  
\nonumber \\& = &  
H_2 \left(\frac {1- \gamma }{2}\right )- H_2 \left(\frac \gamma  2 \right ) 
\,,\label{bounddamp}
\end{eqnarray}
where the optimal vector of probabilities is given by
%\begin{eqnarray}
 $\vec{p}= \left (1 -\gamma  /2 \,,0 \,,0\,, \gamma /2 \right )$,
%\;,  \; 
%\end{eqnarray}
and it corresponds to the basis in Eq. (\ref{b1}), with 
$a=\frac{1+\sqrt {1 - \gamma   }}
{\sqrt{2(2- \gamma )}}$, 
$b=\frac{\gamma }{(1+\sqrt {1 - \gamma   }) \sqrt{2(2- \gamma )}}$, and $c=d=\frac{1}{\sqrt 2}$.  
This basis is clearly made of projectors on the eigenstates of the output 
state (\ref{outdamp}).  
It turns out that, as long as $\gamma < 1/2$, 
a non-vanishing quantum capacity is detected. 
Indeed the difference $Q-Q_{DET}$ never exceeds $0.005$.  
We notice that the Bell basis (\ref{phipsi}) 
does not provide the minimum value of $H(\vec p)$. Actually, 
in such case one has 
\begin{eqnarray}
\vec{p}= \frac 14 \left (
(1 +\sqrt{1- \gamma } )^2 \,,(1  -\sqrt{1- \gamma  })^2 \,,\gamma  \,, \gamma \right )
\;,\label{bellvec}
\end{eqnarray}
and by using this value of $\vec p$ a non-vanishing quantum capacity is detected only for 
$\gamma < 0.3466 $. In Fig. \ref{fig:damp2} we plot the detectable 
bound from Eq. (\ref{bounddamp}) [which is indistinguishable from the quantum capacity (\ref{qcdamp})], 
along with the bound obtained by the probability vector (\ref{bellvec}) pertaining to the Bell projectors,  
versus the damping parameter $\gamma  $. The difference of the curves shows how
the optimisation of $Q_{DET}$ over the bases (\ref{b1}-\ref{b3}) is crucial to 
achieve the optimal bound. 
\par
\begin{figure}[htb]
  \includegraphics[scale=0.7]{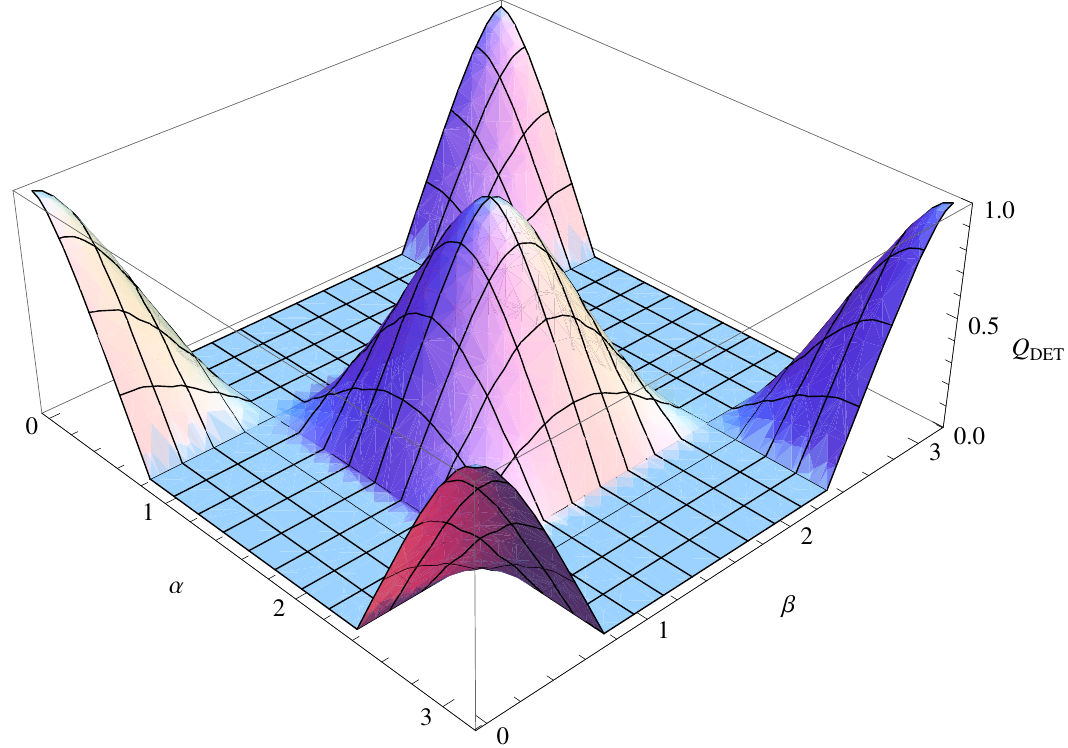}
  \caption{Positive region of the detected quantum channel capacity (\ref{qdetab}) 
for the two-Kraus channel in Eq. (\ref{2k}). }
  \label{fig:2kraus}
\end{figure}

 We finally consider the following set of channels, characterised 
by just two Kraus operators \cite{rsw,wpg,deg},  
\begin{eqnarray}
{\cal E}(\rho )=\sum _{i=1}^2 A_i  \rho A _i^\dag 
\;,\label{2k}
\end{eqnarray}
where $A_1=\cos\alpha |0\rangle\langle 0|
+\cos\beta |1\rangle\langle 1|$ and $A_2=\sin\beta |0\rangle\langle 1|
+\sin\alpha |1\rangle\langle 0|$, with $\alpha,\beta\in\mathbb{R}$. 
%\begin{equation}\label{A2}
%A_1=\left(%
%\begin{array}{cc}
%  \cos\alpha & 0 \\
%  0 & \cos\beta \\
%\end{array}%
%\right),\quad A_2=\left(%
%\begin{array}{cc}
%  0 & \sin\beta \\
%  \sin\alpha & 0 \\
%\end{array}%
%\right)\;.
%\end{equation}
Details on these channels are given in the Supplemental Material \cite{appendix}. 
Our detectable quantum capacity can be written as 
\begin{eqnarray}
Q \geq 
Q_{DET} &= &H_2((\cos ^2 \alpha + \sin ^2 \beta )/2)  \nonumber \\&- &  
H_2 ((\sin ^2 \alpha +\sin ^2 \beta  )/2)
\;.\label{qdetab}
\end{eqnarray}
We checked numerically that $Q - Q_{DET} < 0.005$ for all values of $ \alpha  $ and $ \beta  $. 
The positive region of the 
detected capacity $Q_{DET}$ is plotted in Fig. \ref{fig:2kraus}.

In conclusion, we have proposed a method to detect lower bounds to capacities 
of quantum communication channels, specifically to the quantum capacity, the 
entanglement 
assisted capacity, and the private capacity. The procedure does not require any
a priori knowledge about the quantum channel and relies on a number of measurement 
settings that scales as $d^2$. It is therefore much cheaper than full process
tomography and it can be easily accessed in the lab without posing any 
particular restriction on the nature of the physical system under 
consideration. In particular, for quantum optical systems it is easily 
implementable with present-day technologies \cite{tech}. 
We tested the method for significant qubit channels and it turned out to give
extremely good results for various forms of noise.
The method can be successfully applied also to correlated Pauli and amplitude 
damping channels acting on two qubits \cite{unpu}.

We thank Antonio D'Arrigo for valuable suggestions.

%\clearpage

%\setcounter{page}{1}
\setcounter{equation}{0}
\appendix*
\section{Appendix: Supplemental material}
The following set of channels
\begin{eqnarray}
{\cal E}(\rho )=\sum _{i=1}^2 A_i  \rho A _i^\dag 
\;,\label{2k2}
\end{eqnarray}
where $A_1=\cos\alpha |0\rangle\langle 0|
+\cos\beta |1\rangle\langle 1|$ and $A_2=\sin\beta |0\rangle\langle 1|
+\sin\alpha |1\rangle\langle 0|$, with $\alpha,\beta\in\mathbb{R}$. 
%\begin{equation}\label{A2}
%A_1=\left(%
%\begin{array}{cc}
%  \cos\alpha & 0 \\
%  0 & \cos\beta \\
%\end{array}%
%\right),\quad A_2=\left(%
%\begin{array}{cc}
%  0 & \sin\beta \\
%  \sin\alpha & 0 \\
%\end{array}%
%\right)\;.
%\end{equation}
is characterised by just two Kraus operators \cite{rsw,wpg,deg}, 
and represents the normal form of equivalence classes, since two channels have the same capacity
if they differ merely by unitaries acting on input and output. 
Notice that for $\alpha=\beta$ the channel is dephasing, and for $\beta=0$ 
it is amplitude damping. 
These channels are shown to be degradable \cite{wpg} for 
$\cos(2\alpha)/\cos(2\beta)>0$, hence $Q=Q_1$. 
On the other hand, they are antidegradable for 
$\cos(2\alpha)/\cos(2\beta)\leq 0$, 
thus with $Q=0$. 

\par The coherent information is maximised by diagonal input 
states, and 
in the region  $\cos(2\alpha)/\cos(2\beta)>0$ the quantum capacity is 
given by \cite{wpg}
\begin{eqnarray}\label{Qformula}
Q=\max_{p\in[0,1]} && \;H_2 \big(p \cos^2\alpha +(1-p)
\sin^2\beta\big) \nonumber \\& & 
- H_2 \big(p \sin^2\alpha +(1-p)
\sin^2\beta\big)\;.
\end{eqnarray}
For a detection scheme with the maximally entangled input state 
$|\Phi ^+ \rangle $,  
the output state can be shown to be diagonal on the basis 
in Eq. (\ref{b1}), with 
\begin{eqnarray}
&&a=\frac{\cos \beta -\cos \alpha }{\sqrt{2(\cos^2 \alpha +\cos ^2 \beta )}}, 
\quad  
%\nonumber \\& & 
b=\frac{\cos \alpha  +\cos \beta }{\sqrt{2(\cos^2 \alpha +\cos ^2 \beta )}}, \nonumber \\& & 
c=\frac{\sin \beta -\sin \alpha }{\sqrt{2(\sin ^2 \alpha +\sin ^2 \beta )}}, 
%\nonumber \\& & 
\quad d=\frac{\sin \alpha  + \sin \beta }{\sqrt{2(\sin ^2 \alpha +\sin ^2 \beta )}} 
\nonumber \;, 
\end{eqnarray}
and eigenvalues $\{0, (\cos ^2 \alpha +\cos ^2 \beta  )/2, 0, 
(\sin ^2 \alpha +\sin ^2 \beta  )/2 \}$. 
The optimal vector of probabilities $\vec p$ for the detectable bound 
$Q_{DET}$ corresponds to these eigenvalues. The output entropy of the reduced state 
is then given by 
%\begin{eqnarray}
$S\left ({\cal E}\left (\frac I 2 \right )\right )
= H_2((\cos ^2 \alpha + \sin ^2 \beta )/2)$,
%\;,\end{eqnarray}
hence the detectable quantum capacity can be written as 
\begin{eqnarray}
Q \geq 
Q_{DET} &= &H_2((\cos ^2 \alpha + \sin ^2 \beta )/2)  \nonumber \\&- &  
H_2 ((\sin ^2 \alpha +\sin ^2 \beta  )/2)
\;.\label{qdetab2}
\end{eqnarray}

\end{document}